\documentclass[twocolumn,showpacs,aps,prd,nobibnotes,nofootinbib,floatfix]{revtex4-2}

\usepackage{amsmath}
\usepackage{graphicx,subfigure}
\usepackage[usenames]{color}
\usepackage[normalem]{ulem}
\usepackage{textcomp}
\usepackage{gensymb}
\usepackage{xspace}
\usepackage{verbatim}

\usepackage{soul}

\usepackage[colorlinks=true]{hyperref}
\hypersetup{
  citecolor  =red,
  urlcolor=blue
}
\usepackage{orcidlink}
\newcommand{\rh}{r_{\text{h}}}
\newcommand{\dd}{\text{d}}
\newcommand{\ac}{\alpha_{\text{c}}}
\newcommand{\vef}{V_{\text{eff}}}
\newcommand{\rps}{r_{\text{ps}}}
\newcommand{\rhs}{r_{\text{hs}}}
\newcommand{\rhl}{r_{\text{hl}}}

\begin{document}

\title{Deciphering black hole phase transitions through photon spheres}

\author{Si-Jiang Yang\orcidlink{0000-0002-8179-9365}$^a$$^b$}
\email{yangsj@lzu.edu.cn}
\author{Shan-Ping Wu\orcidlink{0000-0001-6633-1054}$^a$$^b$}
\email{120220908841@lzu.edu.cn}
\author{Shao-Wen Wei\orcidlink{0000-0003-0731-0610}$^a$$^b$}
\email{weishw@lzu.edu.cn}
\author{Yu-Xiao Liu\orcidlink{0000-0002-4117-4176}$^a$$^b$}%
\email{liuyx@lzu.edu.cn, corresponding author}
\affiliation{$^{a}$Lanzhou Center for Theoretical Physics, Key Laboratory of Theoretical Physics of Gansu Province, Key Laboratory of Quantum Theory and Applications of MoE, Gansu Provincial Research Center for Basic Disciplines of Quantum Physics, Lanzhou University, Lanzhou 730000, China\\
$^{b}$Institute of Theoretical Physics $\&$ Research Center of Gravitation, School of Physical Science and Technology, Lanzhou University, Lanzhou 730000, China}
\date{\today}

\begin{abstract}
Black hole thermodynamics is a crucial and foundational aspect of black hole physics, yet its observational verification remains exceptionally challenging. The photon sphere of a black hole, a manifestation of strong gravitational effects, is intrinsically linked to its shadow, which has been directly captured through observations made by the Event Horizon Telescope. Investigating black hole thermodynamics from a gravitational perspective presents an intriguing avenue for research. This paper obtains an analytical formula for the coexistence curve and investigates the relationship between the thermodynamic phase transition and the photon sphere of a black hole with quantum anomaly. It proposes that the photon sphere encodes information about the black hole phase transition, arguing that the change in the photon sphere radius can serve as an order parameter characterizing the black hole's phase transition.
\end{abstract}
\maketitle

\section{Introduction}\label{sec:intro}

Black hole thermodynamics remains an active and significant area of research in general relativity, owing to its profound and fundamental connections among gravitation, thermodynamics, and quantum field theory~\cite{Wald:1999vt}. Black hole thermodynamics implies that black holes not only obey the four laws of thermodynamics~\cite{Bekenstein:1972tm,Bardeen:1973gs}, but also possess entropy and may exhibit an underlying microscopic structure~\cite{Strominger:1997eq,Wei:2015iwa,Wei:2019uqg} with a potential statistical origin~\cite{Bekenstein:1975tw,Maldacena:1996gb,Cheng:2024hxh,Cheng:2024efw,Ali:2024adt}.

Recently, thermodynamics for black holes in anti-de Sitter (AdS) spacetime has attracted a lot of attention. For black holes in AdS spacetime, a rich phase structure emerges, closely tied to the profound implications of the anti-de Sitter/conformal field theory (AdS/CFT) correspondence~\cite{Maldacena:1997re,Aharony:1999ti,Witten:1998qj}. Notably, the Hawking–Page phase transition~\cite{Hawking:1982dh}, which describes a transition between the Schwarzschild-AdS black hole and a thermal gas background, has a dual interpretation as the confinement/deconfinement transition of the quark-gluon plasma in the large $N$ gauge theory~\cite{Witten:1998zw}. For a charged-AdS black hole, it was found that the small-large black hole phase transition is very similar to that of the van der Waals fluid~\cite{Chamblin:1999tk,Chamblin:1999hg}. When the negative cosmological constant is interpreted as thermodynamic pressure~\cite{Kastor:2009wy}, it has been demonstrated that the small-large black hole phase transition in the Reissner–Nordstr\"om AdS black hole not only resembles that of a van der Waals fluid, but also exhibits remarkably similar phase behavior, including identical critical exponents~\cite{Kubiznak:2012wp}. Further investigations of various black holes in AdS spacetime within the framework of extended phase space thermodynamics reveal that their phase structures closely resemble those found in conventional thermodynamic systems. These behaviors have been extensively explored across a wide range of gravitational backgrounds~\cite{Wei:2012ui,Wei:2017icx,Cheng:2016bpx,Cai:2013qga,Wei:2021bwy,Yang:2021ljn,Xu:2022jyp,Wu:2024gqi}, such as reentrant phase transitions~\cite{Gunasekaran:2012dq,Frassino:2014pha}, triple points~\cite{Altamirano:2013uqa,Altamirano:2013ane,Wei:2014hba}, superfluid behaviors~\cite{Hennigar:2016xwd,Bai:2023woh}, among others. For recent reviews on black hole thermodynamic phase transitions, see Refs~\cite{Kubiznak:2016qmn,Mann:2024kzk}. The interpretation of a cosmological constant in the extended phase space has achieved great success from different aspects, including the coupling constant of gauge field~\cite{Hajian:2023bhq}, the extended Iyer-Wald formalism~\cite{Xiao:2023lap}, and higher dimensional origins~\cite{Frassino:2022zaz}.

A black hole is not only a thermodynamic system, but also a strong gravitational system. During the merger of two black holes, gravitational waves are emitted~\cite{LIGOScientific:2016aoc,LIGOScientific:2016sjg}. In particular, the gravitational waves observed during the ringdown phase are intimately connected to the black hole’s quasinormal modes~\cite{Berti:2009kk}. Furthermore, when light passes near a black hole, it can be deflected and gravitationally lensed, resulting in the formation of observable shadows, such as those captured by the Event Horizon Telescope~\cite{EventHorizonTelescope:2019dse,EventHorizonTelescope:2019ggy}. The boundary of a black hole shadow is closely associated with the unstable photon sphere (or light ring)~\cite{Gralla:2019xty}.

It has long been anticipated that the thermodynamic properties of black holes have observable signatures, potentially encoded in quasinormal modes and photon spheres, which may be probed by gravitational wave observatories and the Event Horizon Telescope. Studies have shown that the quasinormal mode frequencies of the Reissner-Nordstr\"om black hole exhibit a spiral-like structure in the complex frequency plane near the second-order phase transition point of Davies~\cite{Jing:2008an}. While the investigation of Berti et al. indicates that such a relation may not be so accurate to describe the phase transition~\cite{Berti:2008xu}. However, further investigation confirmed that the position of divergence of the heat capacities can be reflected in the quasinormal spectrum, which shows the nontrivial relation between the thermodynamic and the quasinormal frequencies of black holes~\cite{He:2008im}. Further research on the quasinormal modes of the Reissner-Nordstr\"om AdS black hole found a dramatic change in the slopes of quasinormal frequencies between small and large black holes near the critical point~\cite{Liu:2014gvf}. Systematic research on other black holes further supports the idea that the quasinormal modes encode information about black hole phase transition~\cite{Mahapatra:2016dae,Chabab:2016cem,Zou:2017juz,Prasia:2016esx,Zhao:2025ecg}.

In addition to being reflected in the quasinormal modes of dynamic perturbations, the thermodynamic phase transitions of black holes can also be revealed through the properties of the photon sphere. Previous research has revealed that plots of thermodynamic quantities (e.g., temperature or pressure) against the photon sphere radius exhibit oscillatory behavior during a first-order phase transition. This behavior, however, disappears at the critical point, indicating a direct link between black hole phase transitions and the photon sphere~\cite{Wei:2017mwc}. Further investigations suggest that the changes in the photon sphere radius may serve as an order parameter for characterizing black hole phase transitions~\cite{Wei:2018aqm}. Similar results have been observed in other black holes~\cite{Xu:2019yub,Du:2022quq,NaveenaKumara:2019nnt,Li:2019dai,Zhang:2019tzi,Kumar:2024sdg}.

The photon sphere of a black hole, a manifestation of strong gravitational effects, is intrinsically linked to its shadow, as directly observed by the Event Horizon Telescope. Investigating black hole thermodynamics from a gravitational perspective remains a compelling subject to research. Inspired by this perspective, we further explore the relationship between black hole thermodynamic phase transitions and the photon sphere in the presence of quantum anomaly. This anomaly intrinsically modifies the scaling behavior and leads to a breakdown of conventional scaling relations, with first-order phase transitions occurring both below and above the critical point.

The outline of the paper is as follows. In Sec.~\ref{Sec:2}, we explore the thermodynamics and phase transition for the static spherical charged black hole with trace anomaly, and obtain the coexistence curve from Maxwell's equal area law. In Sec.~\ref{Sec:phs}, we investigate the photon sphere and explore its relationship with black hole phase transitions. The last section is devoted to the conclusion and discussion.

\section{Thermodynamics of black hole with trace anomaly}\label{Sec:2}

\subsection{Black hole with trace anomaly}

In the presence of strong gravitational fields, quantum effects play a crucial role and must be properly accounted for. Among the various approaches to incorporating these effects, one method involves considering the backreaction induced by the conformal anomaly of quantum field theory in black hole spacetimes~\cite{Cai:2009ua,Cai:2014jea,Hu:2024ldp}. Upon incorporating the conformal anomaly, the Einstein field equation takes the form:
\begin{equation}
    R_{\mu\nu}-\frac{1}{2}Rg_{\mu\nu}+\Lambda g_{\mu\nu}=8\pi\langle T_{\mu\nu}\rangle,
\end{equation}
where $\langle T_{\mu\nu}\rangle$ is the energy momentum tensor associated with the trace anomaly. In four dimensions, the trace anomaly has the form~\cite{Cai:2014jea,Duff:1993wm,Deser:1993yx}
\begin{equation}
   g^{\mu\nu} \langle T_{\mu\nu}\rangle=\beta I_4-\alpha_{\text{c}} E_4,
\end{equation}
where $I_4$ is given by
\begin{equation}
I_4=C_{\mu\nu\rho\sigma}C^{\mu\nu\rho\sigma},
\end{equation}
and $E_4$ is just the Gauss-Bonnet term
\begin{equation}
    E_4=R^2-4R_{\mu\nu}R^{\mu\nu}+R_{\mu\nu\rho\sigma}R^{\mu\nu\rho\sigma}.
\end{equation}
The spherically symmetric black hole with the quantum anomaly corrected solution has been found~\cite{Cai:2009ua,Cai:2014jea}:
\begin{equation}
    \dd s^2=-f(r)\dd t^2+\frac{1}{f(r)}\dd r^2+r^2(\dd \theta^2+\sin^2\theta\dd \phi^2),
\end{equation}
with the metric function
\begin{equation}
    f(r)=1-\frac{r^2}{4\alpha_{\text{c}}}\left(1-\sqrt{1-8\alpha_{\text{c}}\left(\frac{2M}{r^3}-\frac{Q^2}{r^4}-\frac{1}{l^2}   \right)}\right),
\end{equation}
where $M$ and $Q$ are the mass and the charge of the black hole, and $l$ is related to the cosmological constant by $\Lambda=-3/l^2$. The metric reduces to the Reissner-Nordstr\"om AdS spacetime in the limit $\alpha_{\text{c}}=0$. Given that the quantum correction is small, the spacetime is asymptotically Reissner-Nordstr\"om AdS spacetime.

The radius $\rh$ of the black hole event horizon is determined by the equation $f(\rh)=0$.
The thermodynamical quantities of the black hole are~\cite{Hu:2024ldp}
\begin{eqnarray}
 M&=&\frac{-2 \ac  l^2+l^2 Q^2+l^2 \rh^2+\rh^4}{2 l^2 \rh},\\
 T&=&\frac{\rh^2+3\rh^4/l^2-Q^2+2\ac}{4\pi\rh^3-16\pi\ac\rh},\label{Temperature}\\
 S&=&\pi\rh^2-4\pi\ac\ln\left(\frac{4\pi\rh^2}{A_0}\right), \quad  \Phi=\frac{Q}{\rh},
\end{eqnarray}
where $A_0$ is an integration constant that depends on the microscopic details of the quantum gravity theory. For simplicity,  usually, $A_0$ is chosen as $8\pi \ac$~\cite{Cai:2014jea}.

In the extended phase space thermodynamics, the negative cosmological constant is interpreted as thermodynamic pressure and the mass is interpreted as enthalpy~\cite{Kastor:2009wy}. The thermodynamic pressure is
\begin{equation}
    P=-\frac{\Lambda}{8\pi}.
\end{equation}
The thermodynamic volume of the black hole is
\begin{equation}
    V=\left(\frac{\partial M}{\partial P}\right)_{S,Q}=\frac{4}{3}\pi\rh^3.
\end{equation}
From Eq.~\eqref{Temperature}, we can get the equation of state, which is
\begin{equation}
    P=-\frac{\ac}{4\pi \rh^4}+\frac{Q^2}{8\pi\rh^4}+\frac{T}{2\rh}-\frac{1}{8\pi\rh^2}-\frac{2\ac T}{\rh^3}.
\end{equation}
In the extended phase space, the first law of black hole thermodynamics is
\begin{equation}
    dM=TdS+\Phi dQ+VdP.\label{1stlaw}
\end{equation}

\subsection{Critical points and coexistence curve}

In the extended phase space, the thermodynamic behavior of the charged black hole with trace anomaly exhibits a rich and intricate phase structure. As will be shown later, the system displays swallowtail behavior and features two critical points. In this subsection, we focus on analyzing the phase structure and deriving the coexistence curve.  The main new result presented in this subsection is the derivation of the analytical expression for the coexistence curve, which, remarkably, is found to be independent of the black hole charge in the reduced parameter space. While in previous work the coexistence curve was obtained numerically via point-by-point plotting~\cite{Hu:2024ldp}, our work provides a fully analytical form.

The critical point of the phase transition for the black hole with quantum anomaly is determined by
\begin{equation}
    \begin{split}
       \left( \frac{\partial P}{\partial \rh}\right)_{T,Q}=0, \qquad \left( \frac{\partial^2 P}{\partial^2 \rh}\right)_{T,Q}=0.
    \end{split}
\end{equation}
From the above equations, we can get two critical points for $0\leq \ac\leq Q^2/8$. These critical points are given by
\begin{equation}
    \begin{split}
    r_{\text{hc}}&=\sqrt{3Q^2-12 \ac+\Gamma}, \qquad V_{\text{c}}=\frac{4}{3}\pi r_{\text{hc}}^3, \\
    P_{\text{c}}&=\frac{6 Q^2-18 \ac+\Gamma}{24 \pi  r_{\text{hc}}^4}, \qquad\quad T_{\text{c}}=\frac{3 Q^2-\Gamma}{48 \pi  \ac r_{\text{hc}}},\label{cp1}
    \end{split}
\end{equation}
and
\begin{equation}
    \begin{split}
        r'_{\text{hc}}&=\sqrt{3Q^2-12 \ac-\Gamma},\qquad V'_{\text{c}}=\frac{4}{3}\pi r'^3_{\text{hc}},\\
         P'_{\text{c}}&=\frac{6 Q^2-18 \ac-\Gamma}{24 \pi  r'^4_{\text{hc}}}, \qquad\quad T'_{\text{c}}=\frac{3 Q^2+\Gamma}{48 \pi  \ac r'_{\text{hc}}},\label{cp2}
    \end{split}
\end{equation}
where we have defined $\Gamma=\sqrt{192 \ac^2-96 \ac Q^2+9 Q^4}$. For $0 \leq \ac<Q^2/8$, there are two critical points, and the critical exponents satisfy the scaling law. However, when $\ac=Q^2/8$, the two distinct critical points~\eqref{cp1} and~\eqref{cp2} merge into a single one, which is
\begin{equation}
    \begin{split}
        r_{\text{hc}}&=\sqrt{\frac{3}{2}}Q, \qquad V_{\text{c}}=\frac{4}{3}\pi r_{\text{hc}}^3,\\
        P_{\text{c}}&=\frac{5}{72 \pi  Q^2}, \qquad T_{\text{c}}=\frac{1}{\sqrt{6} \pi Q}.
    \end{split}
\end{equation}
In this case, the thermodynamic behavior of black holes with quantum anomalies reveals novel and unexpected phase transitions. The Gibbs free energy exhibits a characteristic swallowtail structure not only below but also above the critical pressure, and critical exponents deviate from the conventional values and violate standard scaling laws. These deviations arise intrinsically from the quantum conformal anomaly, which fundamentally alters the scaling relations and establishes a clear causal link between quantum effects and scaling law violation~\cite{Hu:2024ldp}.
In the following, we consider the case $\ac=Q^2/8$.

In the extended phase space, the Gibbs free energy of the black hole is
\begin{equation}
\begin{split}
      G&=M-TS\\
    &=\frac{32 \pi  P \rh^4+9 Q^2+12 \rh^2}{24\rh}\\
    &-\frac{\left(32 \pi  P \rh^4-3 Q^2+4\rh^2\right) \left(  \rh^2-   Q^2 \log \left(\frac{2 \rh}{Q}\right)\right)}{16\rh^3-8Q^2 \rh}.\label{GbF}
\end{split}
\end{equation}

\begin{figure*}
    \centering
    \subfigure[]{\includegraphics[width=0.38 \textwidth]{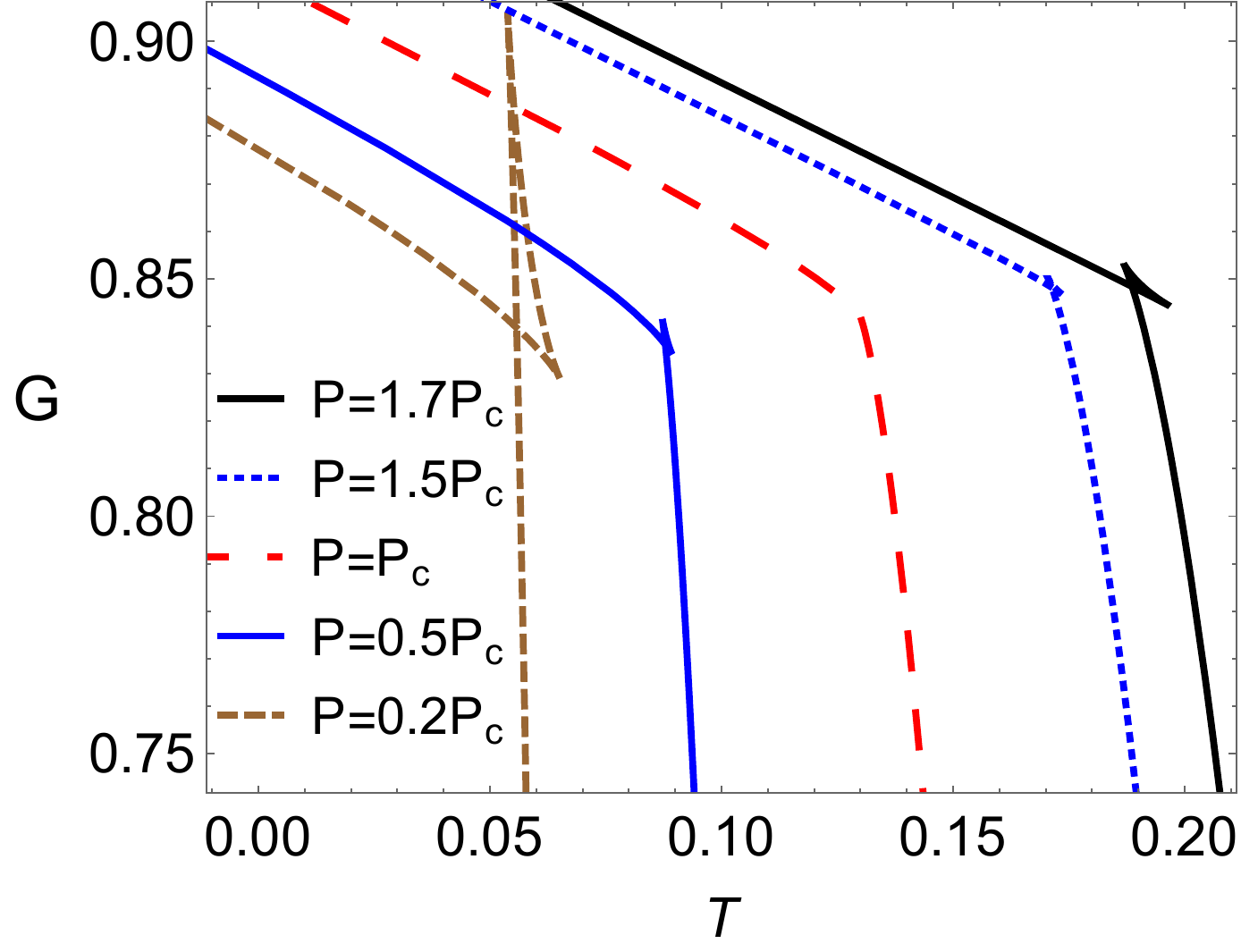}\label{fig:GT}}
    \subfigure[]{\includegraphics[width=0.38 \textwidth]{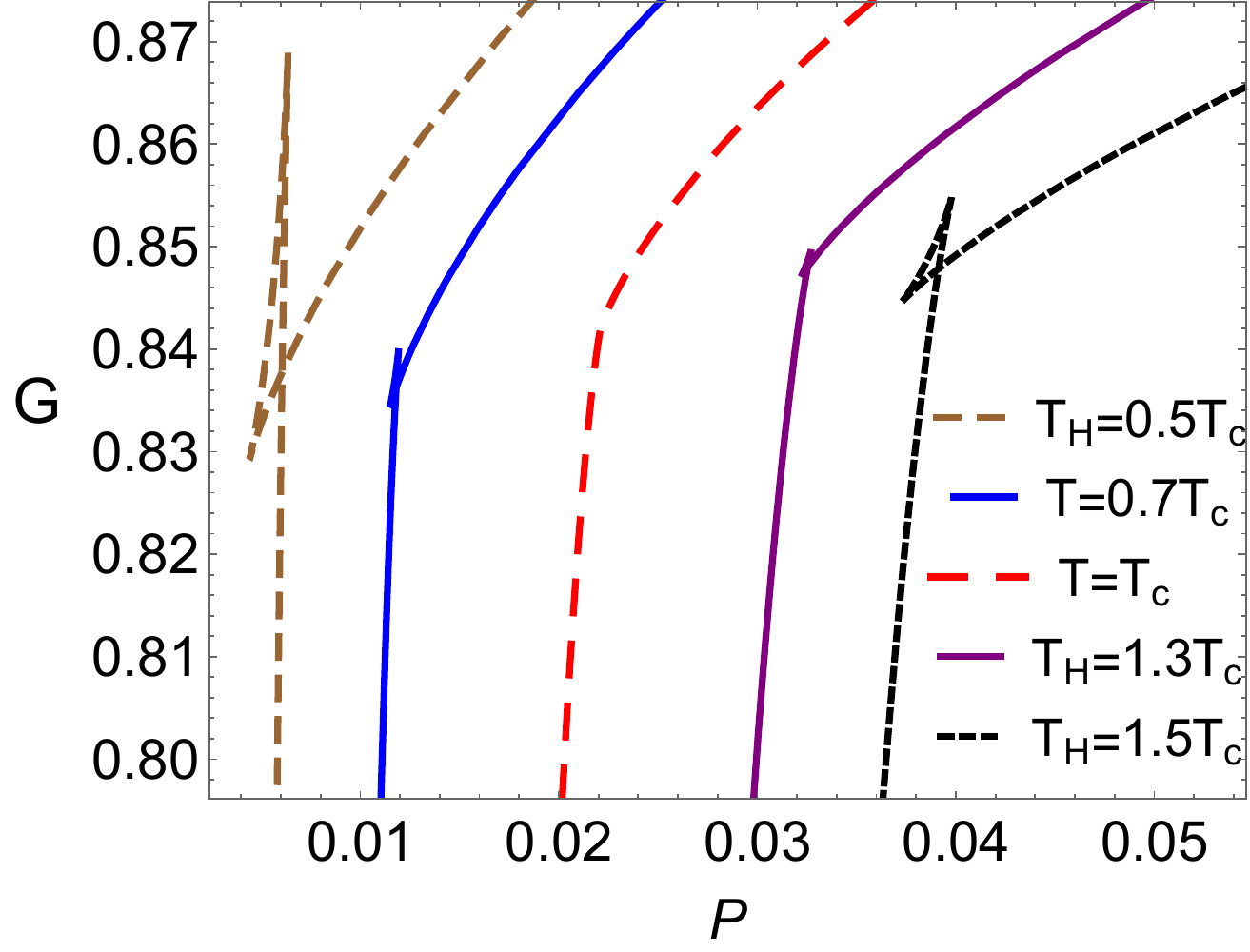}\label{fig:GP}}
    \caption{The Gibbs free energy $G$ vs temperature $T$ and pressure $P$. The Gibbs free energy exhibits swallowtail behavior below and above the critical point. The swallowtail behavior disappears at critical temperature and critical pressure. Here we have set $\ac=Q^2/8$ and $Q=1$.}
    \label{photonsphere}
\end{figure*}

The Gibbs free energy is typically analyzed in an ensemble with fixed pressure $P$ and charge $Q$, and is treated as a function of temperature, as shown in Fig.~\ref{fig:GT}.
 As illustrated in the figure, the Gibbs free energy exhibits a characteristic swallowtail behavior below the critical point, indicating the presence of a first-order phase transition. The swallowtail behavior disappears at the critical point.  In contrast to the phase transition observed in the Reissner-Nordstr\"om AdS black hole, the thermodynamics of a black hole with quantum anomaly not only exhibits swallowtail behavior below the critical temperature, but also shows this behavior at temperatures above the critical point, indicating there exists a first-order phase transition below and above the critical point.

To investigate the coexistence curve in the $P-V$ plane, we consider the Gibbs free energy in the canonical ensemble with fixed temperature $T$ and charge $Q$. We depict the Gibbs free energy $G$ as a function of the thermodynamic pressure $P$ in Fig.~\ref{fig:GP}. As shown in the figure, swallowtail behavior occurs both below and above the critical point but disappears precisely at the critical point.

\begin{figure}
    \centering
    \includegraphics[width=0.4 \textwidth]{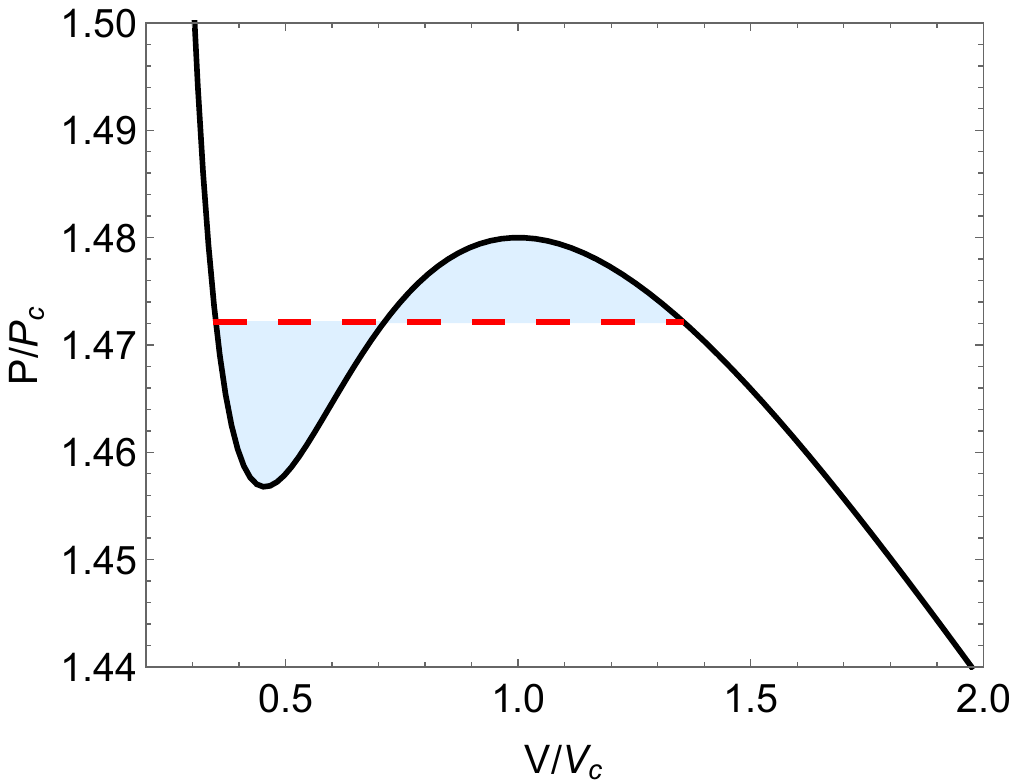}
    \caption{Maxwell's equal area law in the $P-V$ plane. The curve oscillates for a pair of conjugate thermodynamic variables as a first-order phase transition takes place and the areas in the two shadowed regions are the same. Here we have chosen $T=1.3T_{\text{c}}$. }\label{fig:Maxwell}
\end{figure}

At the phase transition point in the canonical ensemble with fixed temperature $T$ and charge $Q$, the Gibbs free energies of the small and large black hole phases become equal. Consequently, the change in Gibbs free energy across the transition satisfies
\begin{equation}
    \Delta G=0.\label{DG}
\end{equation}
From the definition of the Gibbs free energy~\eqref{GbF} and the first law of black hole thermodynamics~\eqref{1stlaw}, we have
\begin{equation}
    dG=-SdT+\Phi dQ+VdP.
\end{equation}
Integrating it and referring to Eq.~\eqref{DG}, we have
\begin{equation}
    \oint \,\dd G=-\oint S\,\dd T+\oint \Phi \,\dd Q+\oint V\,\dd P=0.
\end{equation}
Clearly, the exact result of the integration depends on the physical process. For the fixed $(T,Q)$ canonical ensemble, we have
\begin{equation}
    \oint V\,\dd P=0.\label{MaxwellEA}
\end{equation}
This result shows that the areas of the shadowed regions in Fig.~\ref{fig:Maxwell} are equal. This suggests that Maxwell's equal law holds in the $P-V$ plane.

To get the coexistence curve, we define a dimensionless parameter $\epsilon$~\cite{Wei:2023mxw}
\begin{equation}
    \epsilon=1-\frac{r_{\text{hs}}}{r_{\text{hl}}},
\end{equation}
where $\rhs$ and $\rhl$ are the horizon radii of the saturated coexistence small and large black holes at the same temperature and pressure.

Define the parameters:
\begin{equation}
    \begin{split}
    p=\frac{P}{P_{\text{c}}}, \qquad \tau=\frac{T}{T_{\text{c}}}, \qquad \Tilde{r}_{\text{h}}=\frac{r_{\text{h}}}{r_{\text{hc}}}.\label{eq:redP1}
    \end{split}
\end{equation}
Then, the equation of state for the black hole with a quantum anomaly can be expressed as
\begin{equation}
    p=\frac{12 \Tilde{r}_{\text{h}}^3 \tau -6 \Tilde{r}_{\text{h}}^2-4 \Tilde{r}_{\text{h}} \tau +3}{5 \Tilde{r}_{\text{h}}^4}.\label{eq:eqos}
\end{equation}
Interestingly, the equation of state in the reduced parameter space is independent of the black hole charge $Q$.

From Maxwell's equal area law in the $P-V$ plane, Eq.~\eqref{MaxwellEA}, we have
\begin{align}
    \oint \Tilde{r}_{\text{h}}^3 \,\dd p&=0,\\
    p(\Tilde{r}_{\text{hs}},\tau)&=p(\Tilde{r}_{\text{hl}},\tau).
\end{align}

\begin{figure}
    \centering
    \includegraphics[width=7cm]{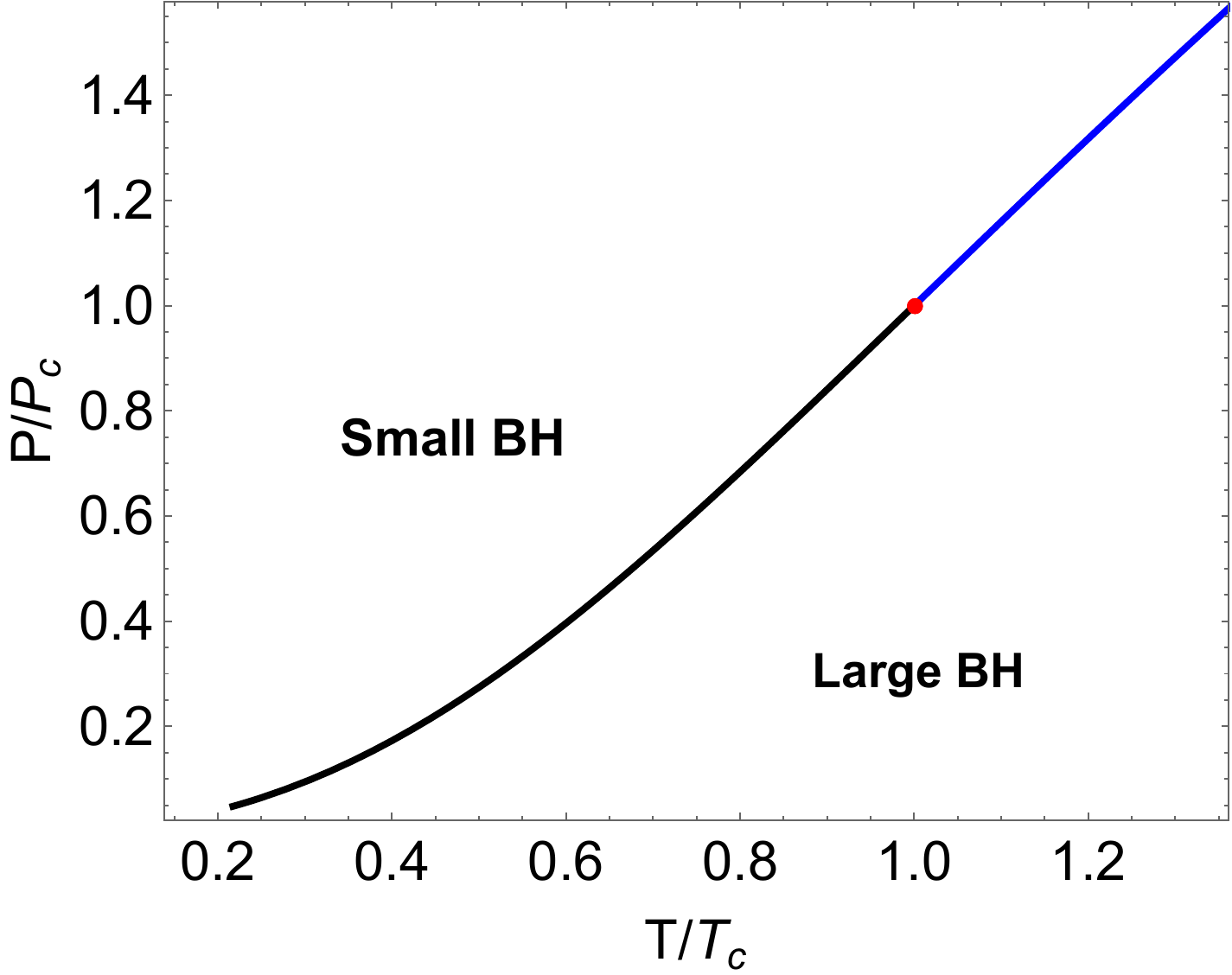}
    \caption{The coexistence curve of phase transition for the black hole with quantum anomaly. In terms of the reduced parameters, the coexistence curve is independent of charge $Q$. The red dot denotes the critical points. In contrast to conventional black hole thermodynamic systems, first-order phase transitions occur both below and above the critical point. }
    \label{fig:coexistC}
\end{figure}

From the above two equations and the definition of the dimensionless parameter $\epsilon$, we can get the analytical coexistence curve in terms of the dimensionless parameter $\epsilon$. Due to the complex form of the formula for the coexistence curve, we just show it in Fig.~\ref{fig:coexistC}.
In the reduced parameter space, the coexistence curve is independent of the black hole charge $Q$. As shown in the figure, small and large black holes coexist below the critical point.
Different from the coexistence curve for Reissner-Nordstr\"om AdS black hole, the small-large black hole coexistence line does not terminate at the critical point. Small-large black hole phase transition also persists above the critical point.

\section{Photon sphere and black hole phase transition}\label{Sec:phs}

On the gravitational side, black holes represent strong-gravity system, with the photon sphere playing a crucial role. In particular, photon sphere is closely related to black hole shadows. In this section, we explore the photon sphere of a black hole with trace anomaly and investigate its potential link to black hole phase transitions, aiming to uncover insights into phase transitions from the photon sphere.

\subsection{Photon sphere for black hole with trace anomaly}

Photon spheres are unstable null circular orbits, and they play a central role in black hole shadows.  In this subsection, we consider the photon sphere of the black hole with trace anomaly.

The motion of photons in the spacetime is governed by null geodesics. The Lagrangian is given by
\begin{equation}
    \mathcal{L}=\frac{1}{2}g_{\mu\nu}\frac{dx^{\mu}}{d\lambda}\frac{dx^{\nu}}{d\lambda}\equiv \frac{1}{2}g_{\mu\nu}\dot{x}^\mu\dot{x}^\nu,
\end{equation}
where $\lambda$ is the affine parameter for the null geodesics.
We consider the free photon moving in the equatorial plane. From the Lagrangian of the free photon, we can get
\begin{equation}
    2\mathcal{L}=-f(r)\dot{t}^2+\frac{1}{f(r)}\dot{r}^2+r^2\dot{\phi}^2.
\end{equation}
The canonical momentum of the photon is
\begin{equation}
    P_{\mu}=\frac{\partial\mathcal{L}}{\partial \dot{x}^\mu}=g_{\mu\nu}\dot{x}^\nu.
\end{equation}
The Lagrangian of the photon is time translation and rotation invariant. Then, the energy and angular momentum of the photon are conserved, which are
\begin{align}
    P_t&=-f(r)\dot{t}\equiv -E=\text{const},\label{energy}\\
    P_{\phi}&=r^2\dot{\phi}\equiv L=\text{const},\label{angularm}
\end{align}
where $E$ and $L$ are the energy and angular momentum of the photon, respectively.

\begin{figure}
    \centering
    \includegraphics[width=7cm]{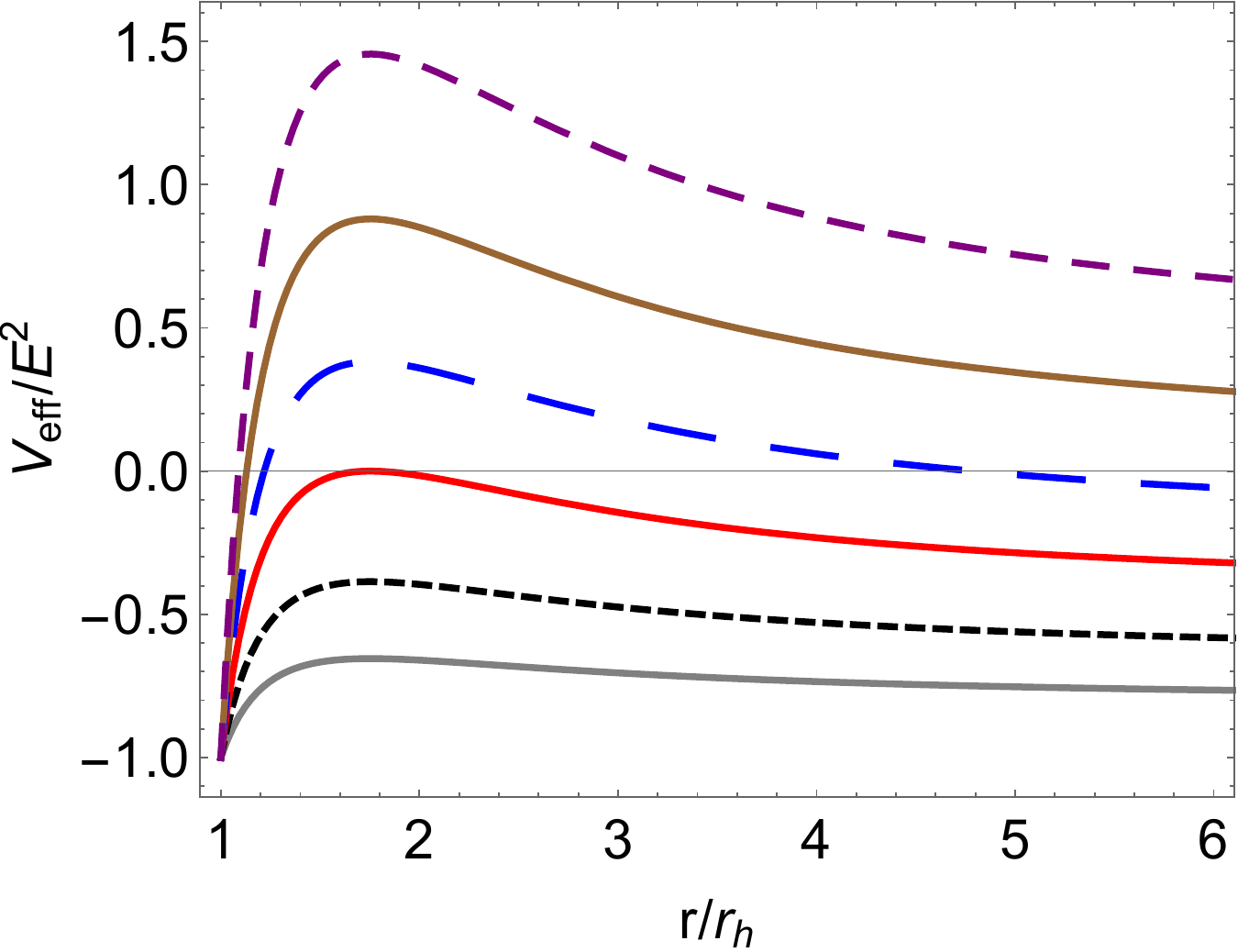}
   \caption{The effective potential for a photon outside the horizon of the four-dimensional black hole with quantum anomaly. The parameters are set to $M=1$, $\ac=Q/8$, $Q=1$, and $P=0.5P_{\text{c}}$. The angular momentum $L/E$ of the photon varies from $1.5$ to $4.0$ from bottom to top. The red thick line corresponds to the effective potential with critical angular momentum $L_{\text{c}}/E=b_{\text{ps}}\approx 2.6$. }
    \label{fig:effV}
\end{figure}

From the Lagrangian for the motion of the photon, we can get the radial equation of motion for the photon, which is
\begin{equation}
    \begin{split}
       2 \mathcal{L}&=-f(r)\dot{t}^2+\frac{\dot{r}^2}{f(r)}+r^2\dot{\phi}^2\\
        &=-\frac{E^2}{f(r)}+\frac{\dot{r}^2}{f(r)}+\frac{L^2}{r^2}=0.
    \end{split}
\end{equation}
From the above equation, the radial equation can be expressed as
\begin{equation}
    \dot{r}^2+V_{\text{eff}}=0,
\end{equation}
where the effective potential is defined as
\begin{equation}
    \vef=\frac{L^2}{r^2}f(r)-E^2.
\end{equation}
Figure~\ref{fig:effV} depicts the effective potential $\vef$ as a function of the radial coordinate $r$ outside the black hole event horizon. The parameters are set to $M=1$, $\ac=Q/8$, $Q=1$, and $P=0.5P_{\text{c}}$ in the figure. The angular momentum $L/E$ of the photon varies from $1.5$ to $4.0$ from bottom to top. The red thick line corresponds to the effective potential with critical angular momentum $L_{\text{c}}/E=b_{\text{ps}}\approx 2.6$.

The circular unstable photon sphere is located at
\begin{equation}
    \begin{split}
        \vef=0, \quad \frac{\partial\vef}{\partial r}=0, \quad \frac{\partial^2\vef}{\partial r^2}<0.
    \end{split}
\end{equation}
From the second equation, we can get the condition for the radius $r_{\text{ps}}$
\begin{equation}
    2f(\rps)-\rps\partial_r f(\rps)=0.\label{eq:phtonR}
\end{equation}

From the formula for the energy~\eqref{energy} and angular momentum~\eqref{angularm} of the photon, we can get the impact parameter
\begin{equation}
 b_{\text{ps}}=\frac{L_{\text{c}}}{E}=\frac{r_{\text{ps}}}{\sqrt{f(r_{\text{ps}})}}.
\end{equation}
The impact parameter, which determines the apparent size of the black hole shadow, is related to its radius through the formula~\cite{Cai:2021fpr}
\begin{equation}\label{rshadow}
  r_{\text{sh}}=r_{\text{ps}}\sqrt{\frac{f(r_0)}{f(r_{\text{ps}})}}=b_{\text{ps}}\sqrt{f(r_0)},
\end{equation}
here, $r_0$ denotes the position of the observer. The result indicates that the radius of the black hole shadow is proportional to the impact parameter associated with the photon sphere. Therefore, for convenience, we use the impact parameter as a measure of the shadow radius.

At the critical point of the black hole phase transition,  the radius of the photon sphere and the corresponding impact parameter are found to be
\begin{align}
     r_{\text{phc}}&=2.35788Q,\\
     b_{\text{phc}}&=2.08213 Q.
\end{align}
Remarkably, the critical impact parameter is smaller than the critical photon sphere radius, which stands in sharp contrast to the case of massive black holes in asymptotically flat spacetimes.

To explore the relationship between the photon sphere and the thermodynamic phase transition, we define the reduced photon sphere radius and reduced impact parameter
\begin{align}
    \Tilde{r}_{\text{ps}}&=\frac{r_{\text{ps}}}{r_{\text{psc}}},\\
    \Tilde{b}_{\text{ps}}&=\frac{b_{\text{ps}}}{b_{\text{psc}}}.
\end{align}
As we will see in the next subsection, the reduced photon sphere radius is independent of the black hole charge.

\subsection{Unstable photon sphere and phase transition}

As illustrated in Figs.~\ref{fig:GT},~\ref{fig:GP} and~\ref{fig:Maxwell}, the Gibbs free energy exhibits a characteristic swallowtail structure, while a pair of the conjugate thermodynamic variables displays oscillatory behavior during the small–large black hole phase transition. At the critical point, both the swallowtail structure and the oscillatory behavior disappear.
In this subsection, we investigate the characteristic behaviors of the photon sphere during black hole phase transitions and examine the potential correlation between these transitions and the properties of the photon sphere.

Given that the photon sphere radius is closely tied to the black hole's size, we analyze how thermodynamic variables depend on the photon sphere radius to identify any novel phenomena during the thermodynamic phase transition.

Starting from Eq.~\eqref{eq:phtonR}, the photon sphere radius $r_{\text{ps}}$ can be obtained in terms of the black hole horizon radius $\rh$, electric charge $Q$, and thermodynamic pressure $P$ through direct calculation. Due to the complexity of the expression, the exact formula is omitted here. Instead, we present only the formal representation, which is
\begin{equation}
    \rps=\rps (\rh, Q, P).
\end{equation}
In terms of the reduced black hole horizon radius $\Tilde{r}_{\text{h}}$ and reduced thermodynamic pressure $p$, as defined in Eq.~\eqref{eq:redP1}, we can express the reduced photon sphere radius $\Tilde{r}_{\text{ps}}$
\begin{equation}
    \Tilde{r}_{\text{ps}}=\Tilde{r}_{\text{ps}} (\Tilde{r}_{\text{h}},p).\label{eq:phsreduced}
\end{equation}
Interestingly, in terms of reduced parameters, the photon sphere radius $\Tilde{r}_{\text{ps}}$ is independent of the black hole charge $Q$.

Similarly, in terms of the reduced parameters, we can get the reduced temperature $\tau$ as a function of reduced pressure $p$ and reduced black hole horizon radius $\Tilde{r}_{\text{h}}$, which is
\begin{equation}
    \tau=\frac{5\Tilde{r}_{\text{h}}^4p+6\Tilde{r}_{\text{h}}^2-3}{12\Tilde{r}_{\text{h}}^3-4\Tilde{r}_{\text{h}}}.\label{temred}
\end{equation}
Equations~\eqref{eq:phsreduced} and~\eqref{temred} give a parametric function for the reduced  photon sphere radius $\Tilde{r}_{\text{ps}}$ .

\begin{figure*}
    \centering
    \subfigure[]{\includegraphics[width=0.38 \textwidth]{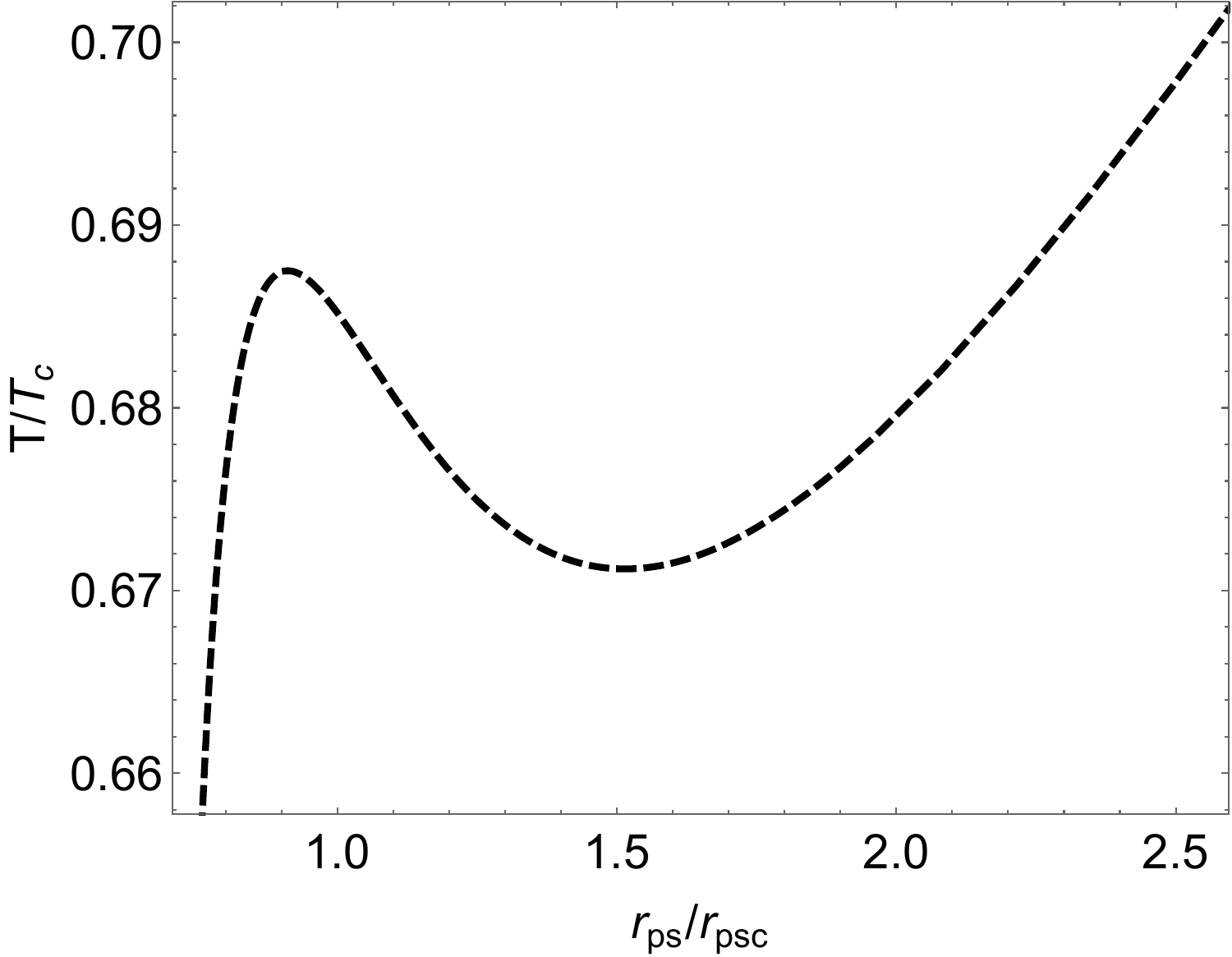}\label{photonspha}}
    \subfigure[]{\includegraphics[width=0.38 \textwidth]{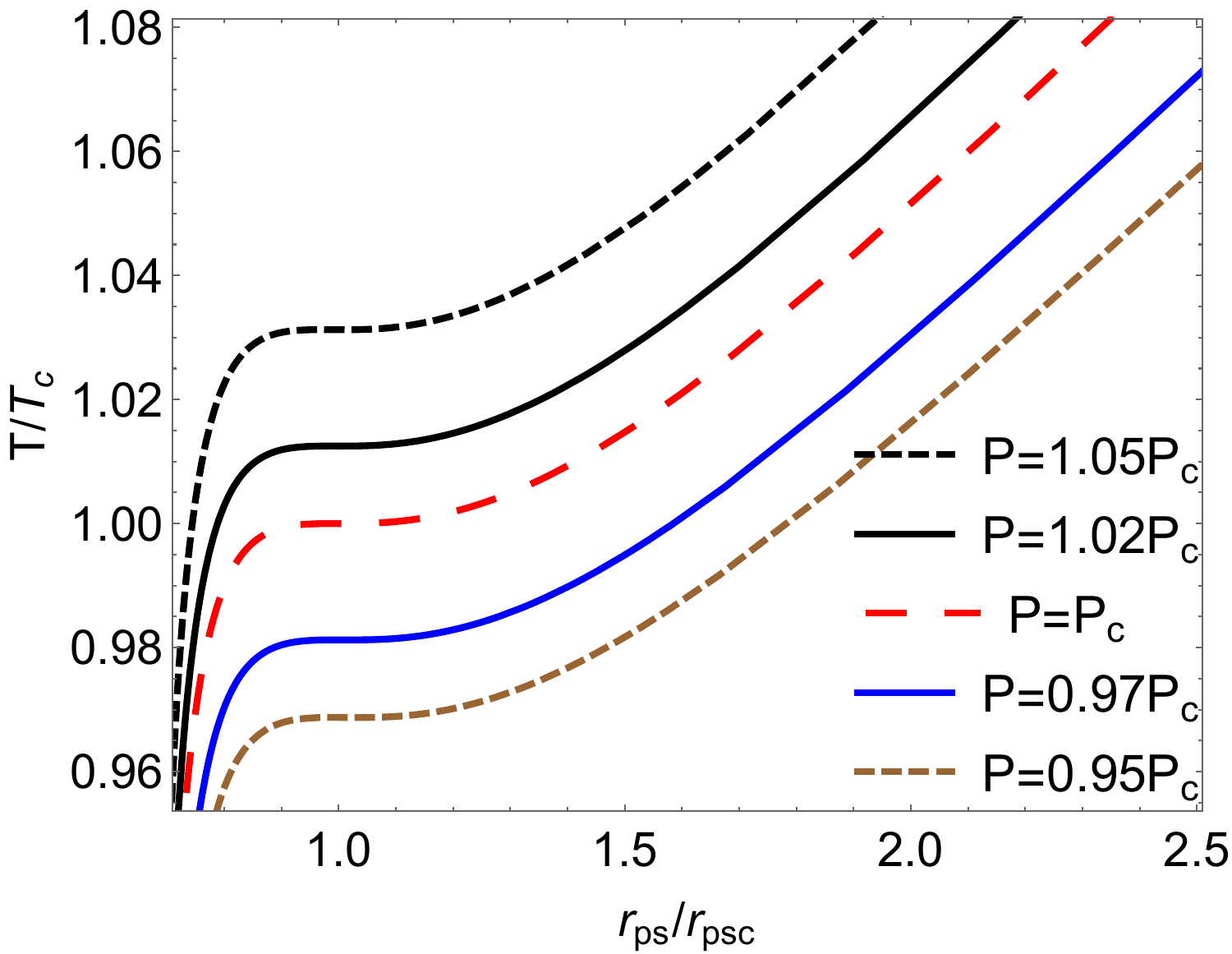}\label{photonsphb}}
    \caption{Phase transition temperature as a function of the reduced photon sphere radius. (a). Phase transition temperature as a function of the reduced photon sphere radius at thermodynamic pressure $P=0.5P_{\text{c}}$. (b). Phase transition temperature as a function of the reduced photon sphere radius at different pressures.}
    \label{photonsphere}
\end{figure*}

To see whether there are novel phenomena for the photon sphere radius during black hole phase transitions, we draw the diagram for the photon sphere radius as a function of the black hole temperature in Fig.~\ref{photonspha}.
\begin{figure*}
    \centering
    \subfigure[]{\includegraphics[width=0.38 \textwidth]{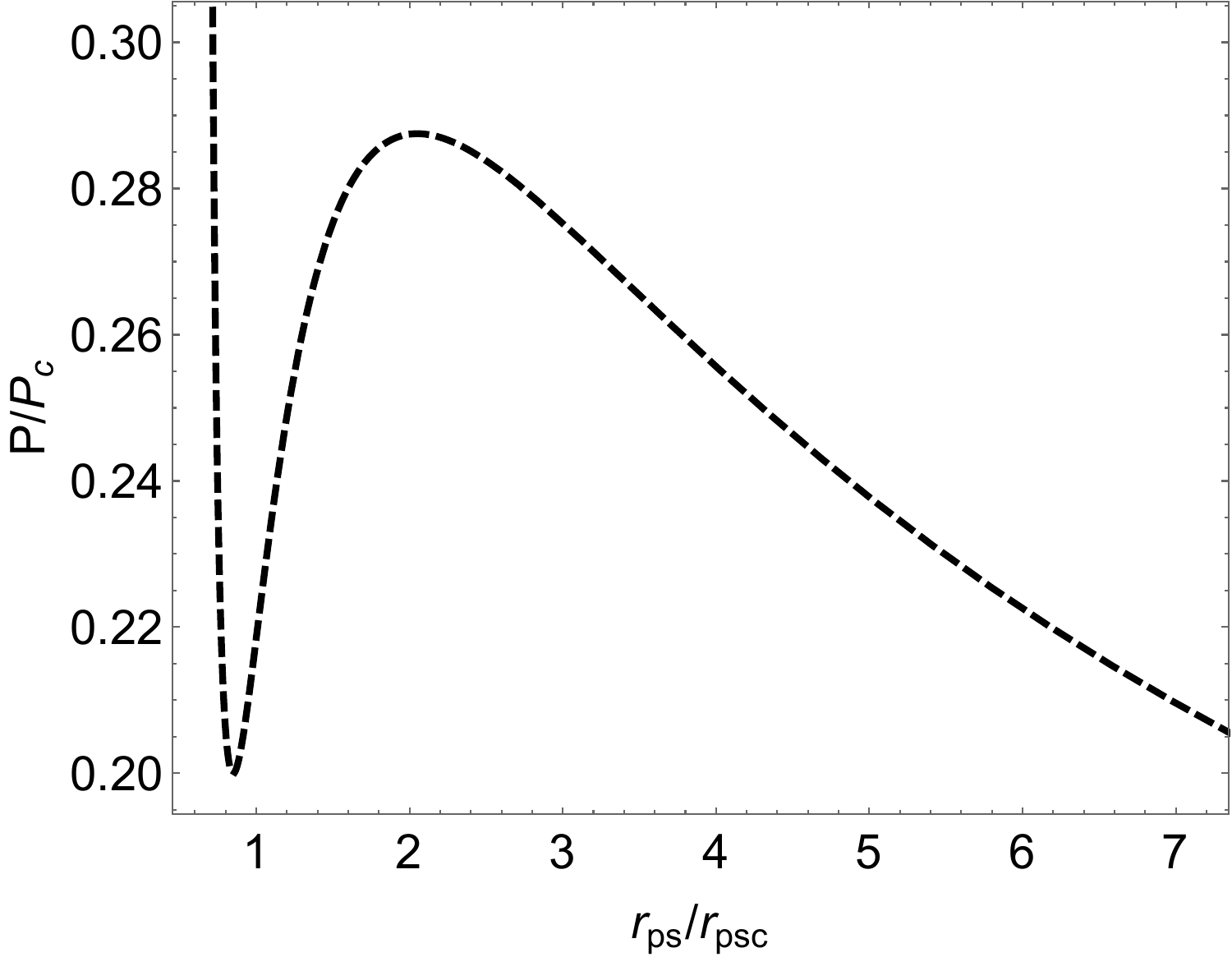}\label{phsa}}
    \subfigure[]{\includegraphics[width=0.38 \textwidth]{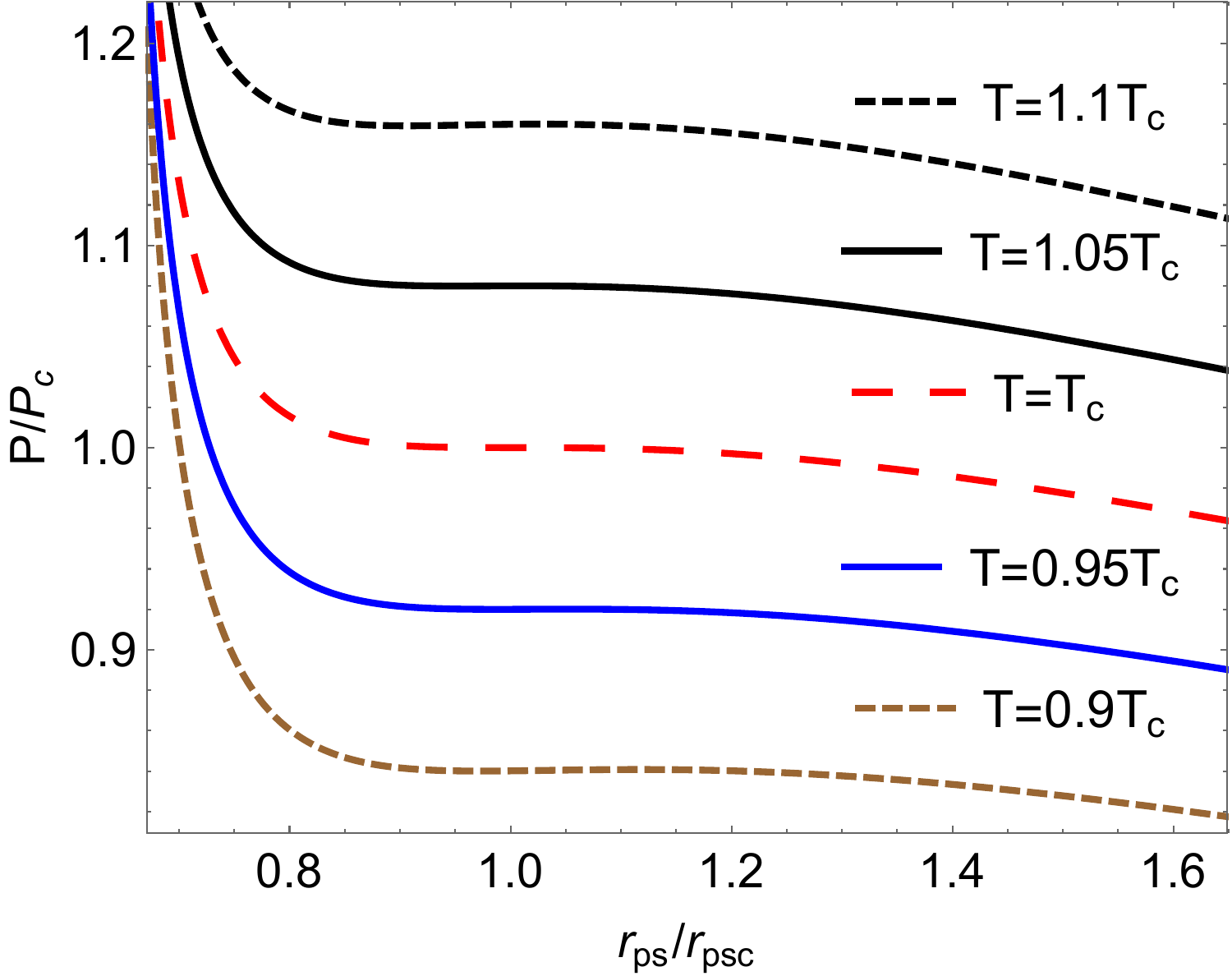}\label{phsb}}
    \caption{Phase transition pressure as a function of the reduced photon sphere radius. (a). Phase transition pressure as a function of photon sphere radius at black hole temperature $T=0.5T_{\text{c}}$. (b). Phase transition pressure as a function of the reduced photon sphere radius at different temperatures.}
    \label{photonsphere}
\end{figure*}
In the figure, we have chosen the thermodynamic pressure $P=0.5P_{\text{c}}$. From the figure, we can see that it oscillates and there are two extremal points. To determine whether this behavior represents a special case or a universal feature of black hole phase transitions, we present a plot for various thermodynamic pressures in Fig.~\ref{photonsphb}. As shown in the figure, below the critical pressure, a first-order phase transition occurs, accompanied by an oscillatory behavior in the $T/T_{\text{c}}-r_{\text{ps}}/r_{\text{psc}}$ curve. At the critical pressure $p=1$, the oscillatory behavior disappears. When the thermodynamic pressure is larger than the critical pressure, there is a first-order phase transition and the oscillatory behavior in the $T/T_{\text{c}}-r_{\text{ps}}/r_{\text{psc}}$ plane appears again. In Fig.~\ref{photonsphb}, the thermodynamic pressure increases from the bottom to the top.

Similarly, the parametric expressions for the reduced photon sphere radius $\Tilde{r}_{\text{ps}}$ and the thermodynamic pressure $p$ can be derived as follows:
\begin{equation}
  \left\{  \begin{split}
    \Tilde{r}_{\text{ps}}&=\Tilde{r}_{\text{ps}} (\Tilde{r}_{\text{h}},\tau),\\
    p&=\frac{12 \Tilde{r}_{\text{h}}^3 \tau -6 \Tilde{r}_{\text{h}}^2-4 \Tilde{r}_{\text{h}} \tau +3}{5 \Tilde{r}_{\text{h}}^4}.
    \end{split}\right.
\end{equation}
From Figs.~\ref{phsa} and~\ref{phsb}, we can see that there are oscillatory behaviors and there exist two extremal points in the $p-\Tilde{r}_{\text{phs}}$ curve for the first-order phase transition, and the oscillatory behavior disappears at the critical point.

This implies that there are hints of black hole phase transitions in the photon sphere structure. We explore it further from the order parameter perspective in the next subsection.

\subsection{Order parameter and critical exponent}

Order parameter is an important quantity to characterize phase transition phenomena.  It changes discontinuously in a first-order phase transition but continuously at the critical point. The order parameter goes to zero at the critical point along the coexistence curve. In this subsection, we explore whether we can use quantities related to the photon sphere as an order parameter to depict black hole phase transitions.

\begin{figure*}
    \centering
    \subfigure[]{\includegraphics[width=0.39 \textwidth]{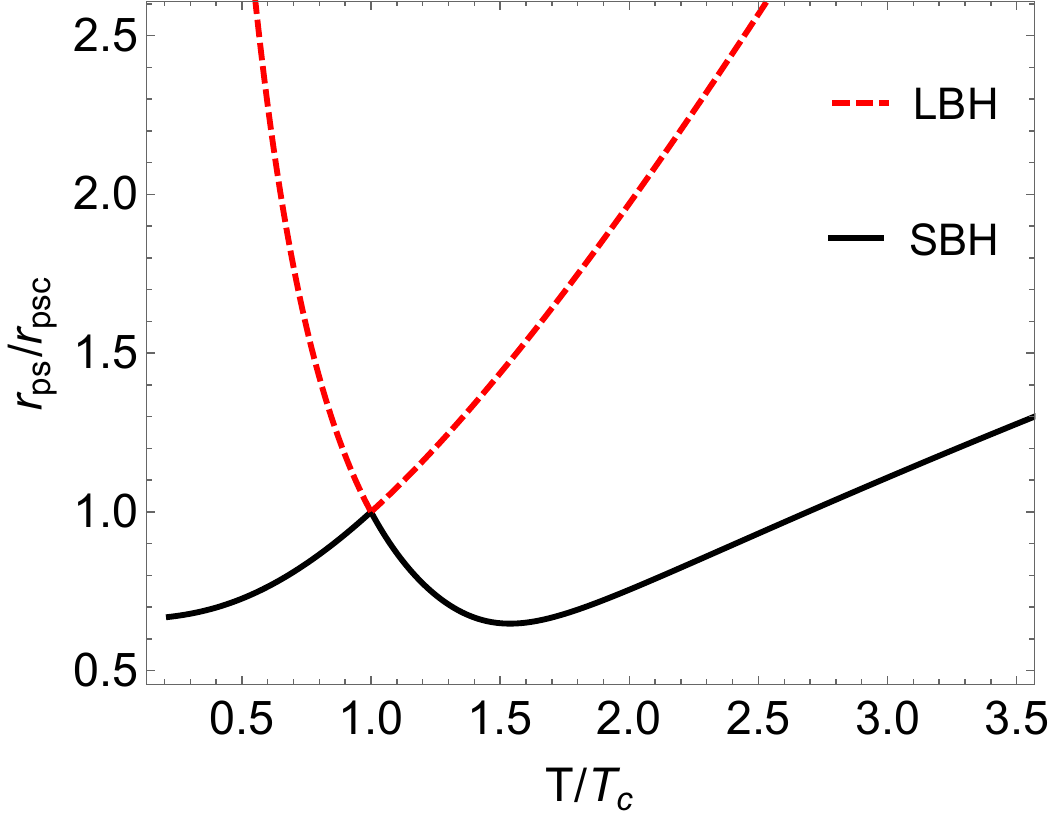}\label{phsradsl}}
    \subfigure[]{\includegraphics[width=0.38 \textwidth]{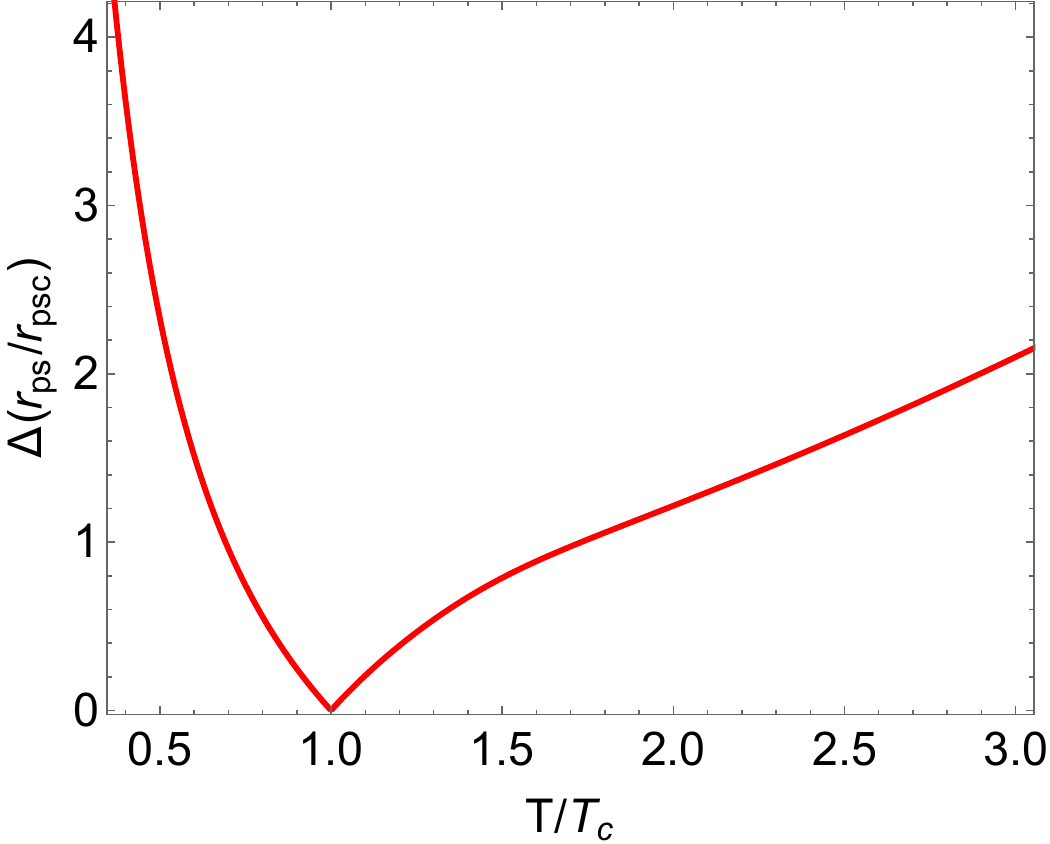}\label{orderP}}
    \caption{(a). The reduced radii of the photon spheres for the coexistence small and large black holes. (b). The difference of the reduced radii of the photon spheres for the coexistence large and small black holes.}
    \label{photonsphere}
\end{figure*}

\begin{figure*}
    \centering
    \subfigure[]{\includegraphics[width=0.39 \textwidth]{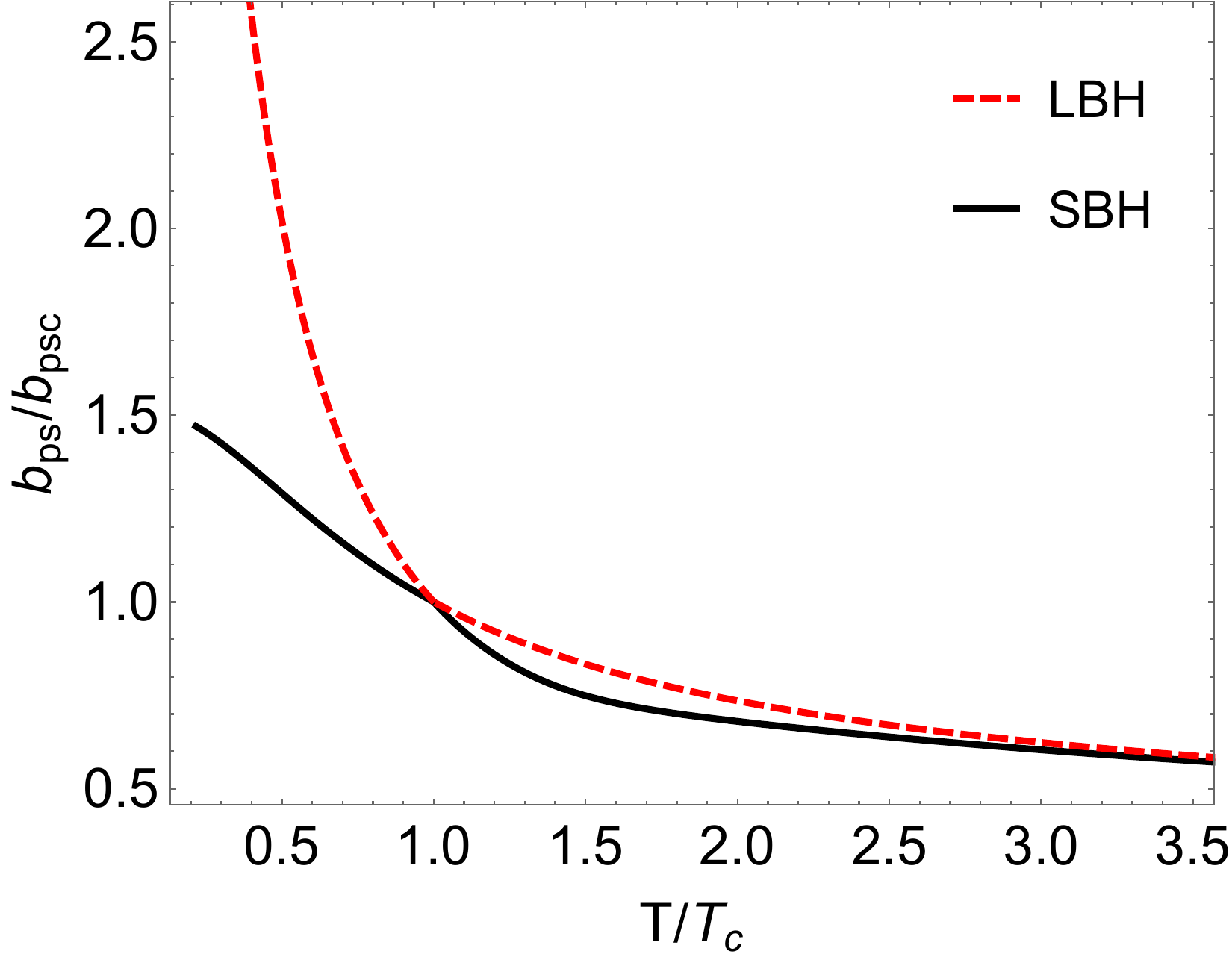}\label{bphsradsl}}
    \subfigure[]{\includegraphics[width=0.38 \textwidth]{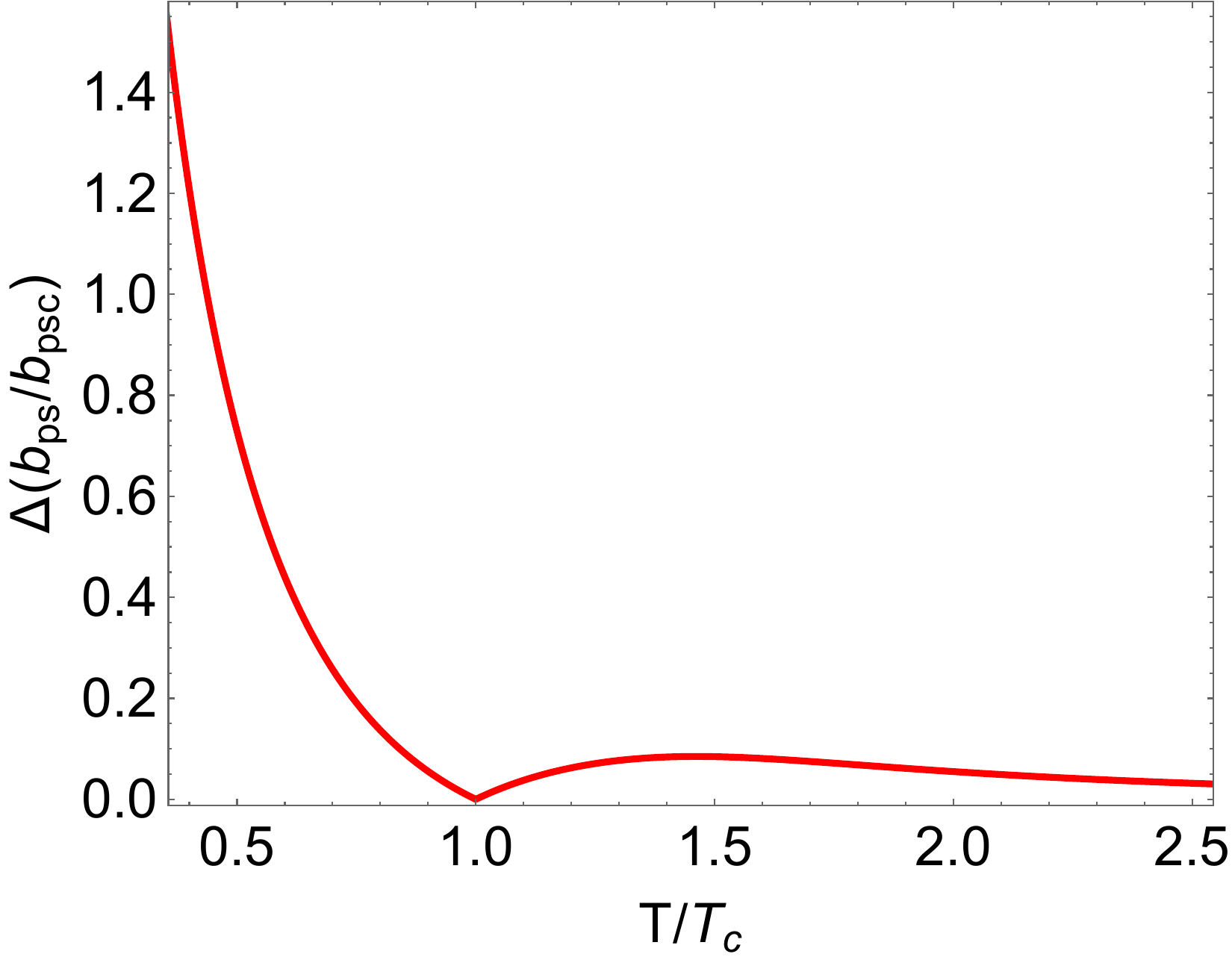}\label{borderP}}
    \caption{(a). The reduced impact parameter for the coexistence small and large black holes. (b). The difference of the reduced impact parameter for the coexistence large and small black holes.}
    \label{impactb}
\end{figure*}

The reduced photon sphere radius, $r_{\text{ps}}/r_{\text{psc}}$ and the reduced impact parameter $b_{\text{ps}}/b_{\text{psc}}$ , are plotted as a function of the phase transition temperature, $T/T_{\text{c}}$, in Figs.~\ref{phsradsl} and~\ref{bphsradsl}. In the reduced parameter space, the reduced photon radius $\Tilde{r}_{\text{ps}}$, the reduced impact parameter $\Tilde{b}_{\text{ps}}$, and phase transition temperature $\tau$ are found to be independent of charge. As illustrated in the figure, below the critical point, the photon sphere radius $\Tilde{r}_{\text{ps}}$ increases for the saturated coexistence small black hole, whereas it decreases for the saturated coexistence large black hole as the temperature increases. The reduced photon radii $\Tilde{r}_{\text{ps}}$ for the coexistence small and large black holes are the same at the critical point $\tau=1$. As the temperature increases, the reduced photon radius $\Tilde{r}_{\text{ps}}$ increases for the coexisting large black hole, whereas it initially decreases and then increases for the coexisting small black hole.

We also plot the difference of the reduced photon sphere radius and the difference of the reduced impact parameter for the saturated coexistence large and small black holes in Figs.~\ref{orderP} and~\ref{borderP}.
From the figure, we can see that the difference of the reduced photon radius decreases with the transition temperature and goes to zero at the critical point. In contrast to the behavior observed in the Reissner-Nordstr\"om AdS black hole~\cite{Wei:2017mwc}, where the difference of the photon sphere radius for the coexistence of large and small black holes terminates at the critical point, the difference of the photon sphere radii for the coexistence of large black holes and small black holes increases with the transition temperature when the temperature exceeds the critical temperature.

The behavior of the difference of the reduced photon sphere radius $\Delta\Tilde{r}_{\text{ps}}$ is extremely similar to the behavior of the order parameter. It exhibits nonzero values during the first-order phase transition and vanishes at the critical point. But we still need to check the critical exponent of the parameter. By performing a series expansion of the reduced photon sphere radius near the critical point, we obtain an analytical expression
\begin{equation}
\begin{split}
    \Delta \Tilde{r}_{\text{ps}}\simeq \left(\frac{\partial \Tilde{r}_{\text{ps}}}{\partial \rh}\right)_{\text{c}}\left(\rhl-\rhs\right)\simeq 2.25\mid \tau-1 \mid.
\end{split}
\end{equation}
The result shows that the critical exponent is $\beta=1$, a value consistent with the result obtained in Ref.~\cite{Hu:2024ldp}.

The impact parameter is directly related to the radius of black hole shadows. Similarly, near the critical point, the difference of the reduced impact parameter for large and small black hole phases can be expressed as
\begin{equation}
\begin{split}
    \Delta \Tilde{b}_{\text{ps}}\simeq \left(\frac{\partial \Tilde{b}_{\text{ps}}}{\partial \rh}\right)_{\text{c}}\left(\rhl-\rhs\right)\simeq 0.46\mid \tau-1 \mid.
\end{split}
\end{equation}
The critical exponent for the reduced impact parameter is still $\beta=1$.

From the definition of the order parameter, it is clear that the changes of the reduced photon sphere radius and the reduced impact parameter for saturated coexistence large and small black holes can be regarded as the order parameter. As suggested in Ref.~\cite{Hu:2024ldp}, the presence of a quantum anomaly leads to a critical exponent of $\beta=1$ for the order parameter, which markedly differs from the value $\beta=1/2$ observed in the Reissner-Nordstr\"om AdS black hole and in most other known black hole cases. Furthermore, the quantum anomaly causes the critical exponents to violate the scaling laws predicted by mean field theory~\cite{Hu:2024ldp}.

\section{Conclusion and discussion}

In this study, we examined the relationship between the thermodynamic phase transition and the photon sphere of a static spherically symmetric charged black hole with trace anomaly. The charged black hole with trace anomaly can be regarded as a quantum correction of the Reissner-Nordstr\"om AdS black hole by incorporating the backreaction induced by the quantum conformal anomaly of the quantum field. Unlike the Reissner-Nordstr\"om AdS black hole, its thermodynamic behavior exhibits two distinct critical points. We considered the special case $\ac=Q^2/8$, where the two critical points merge, and the resulting critical exponents deviate from the scaling laws predicted by mean field theory. In the extended phase space, below and above the critical point, the Gibbs free energy exhibits the characteristic swallowtail behavior indicating a first-order phase transition. At the critical point, the swallowtail behavior disappears. Below and above the critical temperature, there exists an oscillatory behavior in the $P-V$ plane. The oscillatory behavior disappears at the critical temperature. From the oscillatory behavior in the $P-V$ plane, we found that Maxwell's equal area law holds, and we obtained the analytical coexistence curve from Maxwell's equal area law. Similar to the case of the Reissner-Nordstr\"om AdS black hole, the coexistence curve is independent of the electric charge in the reduced parameter space.

After having explored the main thermodynamic features for the charged black hole with trace anomaly, we explored the relationship between the thermodynamical phase transition and the photon sphere.  We found that the thermodynamic variables such as the black hole temperature and thermodynamic pressure oscillate with the unstable photon sphere radius and there are two extreme points when there is a first-order phase transition, and the oscillatory behavior disappears at the critical point. This phenomenon strongly hints that there is a relationship between the unstable photon sphere and black hole phase transition. We further explored the changes of the reduced photon sphere radius for the coexistence large and small black holes; we found it goes to zero at the critical point and changes discontinuously in a first-order phase transition. Furthermore, the critical exponent for the changes of the photon sphere radius for the coexistence large and small black holes is $\beta=1$, which is the same as the order parameter obtained in Ref.~\cite{Hu:2024ldp}. This suggests that the changes of the reduced photon sphere radius can be regarded as the order parameter for the small-large black hole phase transition. In the reduced parameter phase space, all the results for the relationship between phase transitions and the photon sphere are obtained analytically and they are independent of black hole charge.

For the Reissner-Nordstr\"om AdS black hole, the coexistence curve for small-large black holes terminates at the critical point.
An early work showed that the change of the photon sphere radius for $d$-dimensional Reissner-Nordstr\"om AdS black holes can serve as an order parameter for black hole phase transition~\cite{Wei:2017mwc}. When the backreaction of the quantum conformal anomaly is taken into account, the metric of the Reissner-Nordstr\"om AdS black hole is modified and the coexistence curve for small-large black holes does not terminate at the critical point.  We further confirmed that the changes of the photon sphere radius for black holes with trace anomaly can also be regarded as the order parameter even though the coexistence curve does not terminate at the critical point.

The photon sphere encodes geometric and optical information~\cite{Virbhadra:1999nm,Claudel:2000yi}, which is directly linked to observable features such as the black hole shadow. By choosing the photon sphere radius as the order parameter, our aim is to bridge thermodynamic phase transitions and astrophysical observables in a more direct and meaningful way. In this work, we extend the study to phase transitions involving isolated critical points, and further demonstrate that the photon sphere can effectively encode critical behavior. Our results highlight how such thermodynamic phenomena may be imprinted in the photon sphere structure, which could potentially be tested observationally.

As astrophysical black holes are neutron and rotating, for a rotating AdS black hole, thermodynamic information might be effectively encoded in the photon ring, which corresponds to unstable null geodesics confined to the equatorial plane. The properties of the photon ring are determined by the horizon radius, angular momentum, and pressure. Interestingly, it exhibits behavior analogous to that of the photon sphere in spherically symmetric AdS black holes~\cite{Wei:2018aqm}, suggesting a broader universality. In this work, we explored whether such correspondence persists in the presence of quantum anomaly, which intrinsically modifies the scaling relations and leads to the violation of conventional scaling law.
Together with investigations for the relationship between the photon sphere and phase transitions for other black holes~\cite{Wei:2018aqm,Wei:2017mwc,Xu:2019yub}, our work further enhances that the photon sphere might encode information about black hole phase transitions.

Although our universe is asymptotically de Sitter, much of the research on thermodynamics for black holes is carried out in asymptotically AdS. The primary reason lies in the well-defined thermodynamic variables and rich phase behaviors for black holes in AdS space. In AdS space, the gravitational potential of AdS space acts like a confining box, allowing black holes to reach thermal equilibrium~\cite{Hawking:1982dh}.  Moreover, asymptotically AdS black holes provide a gauge dual description of finite-temperature CFTs through the AdS/CFT correspondence~\cite{Maldacena:1997re}. However, black hole thermodynamics in dS space is more intricate due to two factors: the presence of both cosmological and event horizons places the system between them in a nonequilibrium state, and the lack of a globally timelike Killing vector hinders the definition of asymptotic mass~\cite{Kubiznak:2016qmn}. Our exploration of AdS black hole thermodynamics through the photon sphere may shed light on future studies of de Sitter black hole thermodynamics.

These investigations provide a promising avenue for exploring thermodynamic effects from a gravitational perspective, indicating that black hole shadows may encode valuable thermodynamic information. In addition to shadows~\cite{Ladino:2024ned,Ladino:2025oeq}, other observable phenomena—such as particle motion and chaotic dynamics—may also serve as effective probes of black hole phase transitions and might be worth further investigation~\cite{Guo:2022kio,Chen:2025xqc,Hazarika:2025zgv}.

\acknowledgments

The authors thank Yu-Sen An, Wen-Di Guo and Yu-Peng Zhang for helpful discussions. This work was supported by the National Natural Science Foundation of China (Grants No. 12305065, No. 12475056,  No. 12475055 and No. 12247101), the China Postdoctoral Science Foundation (Grant No. 2023M731468), the Gansu Province's Top Leading Talent Support Plan, the Fundamental Research Funds for the Central Universities (Grant No. lzujbky-2024-jdzx06), the Natural Science Foundation of Gansu Province (No. 22JR5RA389), and the `111 Center' under Grant No. B20063.


%
\end{document}